\documentclass[%
 reprint,
showpacs,preprintnumbers,
 amsmath,amssymb,
 aps,
]{revtex4-1}

\usepackage{graphicx}
\usepackage{dcolumn}
\usepackage{bm}

\usepackage[pagebackref=false,colorlinks,linkcolor=blue,filecolor=green,urlcolor=blue ,citecolor=blue]{hyperref}

\usepackage{xcolor}

\usepackage{color,soul} 

\begin{document}

\title{Electron-electron interactions in partially mixed helical states }
\author{Zeinab Bakhshipour and Mir Vahid Hosseini}
 \email[Corresponding author: ]{mv.hosseini@znu.ac.ir}
\affiliation{Department of Physics, Faculty of Science, University of Zanjan, Zanjan 45371-38791, Iran}

\begin{abstract}
We theoretically study the effect of electron-electron interactions in one-dimensional partially mixed helical states. These helical states can be realized at the edges of two-dimensional topological insulators with partially broken time-reversal symmetry, resulting in helical gapped states. Using the bosonization method and renormalization group analysis, we identify weak gap, crossover, and strong gap regimes in the phase diagram. We find that strong electron-electron interaction mixes the helicity of the states, leading to the relevant strong gap regime. We investigate the charge and spin density wave correlation functions in different relevancy regimes of the gap mediated by interactions, where in the case of strong repulsive interaction, the spin density wave dominates the charge density wave. Additionally, employing the Memory function technique, we calculate the effect of mixed helicity on the charge transport in a sufficiently long edge. We find a non-uniform temperature dependence for the charge conductivity in both the strong and weak gap regimes with distinct features. 
\end{abstract}

\maketitle

\section {Introduction} \label{s1}

Topological phase of matters has attracted a lot of attentions from both theoretical and experimental aspects \cite{TI,Ts1,Topo-review3,Ts2}, owing to its promising potential for next-generation technologies \cite{TopoCompu}. In topological insulators \cite{TI}, the spin-Hall effect \cite{SpinHall,QSH_review} is primarily facilitated by spin-orbit (SO) coupling \cite{SO,SOC-review1,SOC-review2}. In three-dimensional time-reversal symmetry-protected topological insulators \cite{TIBiSe,TISbTe}, surface states offer a two-dimensional conducting surface, while the bulk of the material remains in a normal insulating phase, provided the corresponding symmetry is maintained. Additionally, in two-dimensional HgTe quantum wells \cite{BHZTI}, strong SO coupling gives rise to topologically protected edge states, forming one-dimensional (1D) helical conducting channels \cite{spinHallEffect1,spinHallEffect2}. The spin-orbit locking observed in helical systems allows for spin manipulation by altering electron momentum, driving interest in spintronic device fabrication \cite{Spintronics-review} and processing \cite{1Sooriintro}. Furthermore, 1D helical nanowires \cite{SOnanowire1,SOnanowire2} can host topologically nontrivial phases \cite{1DTopoPhase1,1DTopoPhase2,1DTopoPhase3,1DTopoPhase4,1DTopoPhase5,1DTopoPhase6}.

Due to ubiquitous imperfections and impurities, it is challenging to attain an ideal shape for the band structure of materials \cite{BandEngine}. However, this imperfection can serve as the basis for engineering the band structure. For instance, the magnetic impurity doping of topological states is of particular interest, leading to the formation of magnetic topological insulators \cite{DopgapSize1}. In such cases, the moments of magnetic dopants can be ordered ferromagnetically \cite{Cr-doped1,Cr-doped2,Cr-doped3}, resulting in the gapping out of surface states \cite{Masive1,Masive2,Mn-doped,1DGapFormImpu} with magnetically polarized features \cite{PlorizedTI}, while leaving the bulk states unaffected \cite{surfOrder}. Additionally, an effective magnetization, mixing the helicity of the surface states \cite{3Soori2012}, can be established in topological insulators via the proximity effect through coupling to ferromagnetic materials \cite{FeroHetro,TM-REHetro1,TM-REHetro2}. This can provide a 1D gapped Dirac dispersion at the edge of 2D topological states of matter \cite{1DD,BHZTI}.

On the other hand, low dimensionality strongly affects the physical properties of the systems \cite{1D-review0,1D-review1,Giamarchi,1D-review2,1D-review3}. The features of 1D systems have attracted much interest ranging from condensed matter systems to ultracold gases \cite{1DBOS,15someexperi-Giam}. In 1D conducting channels, due to confinement in two transverse directions, the dynamics of a carrier depends essentially on that of the other ones and possesses unique characteristics, namely, bosonic excitations \cite{Tomonaga,Luttinger}. As such, the Fermi liquid theory should be replaced by the Tomonaga-Luttinger liquid theory \cite{TLL}. Subsequently, in the presence of such confining potentials, electron-electron interactions become important resulting in charge-spin separation and the anomalous scaling of correlation functions \cite{ChargSpinSeperaExper}. In addition to the bare interacting 1D systems \cite{Bareee},  multicomponent Tomonaga-Luttinger model \cite{multiee} has aslo been studied. 

Furthermore, the effect of electron-electron interaction in SO-coupled 1D systems has been investigated extensively \cite{eeSOC1D1,eeSOC1D2,eeSOC1D3,eeSOC1D4,eeSOC1D5,13Hsu-chargetransport} within two-loop \cite{eeSOC1D6} or one-loop \cite{eeSOC1D7,eeSOC1D8} renormalization-group (RG) scheme. When SO is strong enough, it provides a divergence of susceptibility near zero temperature and changes the dominant fluctuations \cite{eeSOC1D5}. Furthermore, for 1D interacting systems even an arbitrary weak SO coupling can reveal a strongly correlated phase with a spin gap in the presence of inversion symmetry \cite{symm-protectspingap}. In Rashba nanowires, a helical gap can be emerged by the electron-electron interaction, including spin-umklapp scattering \cite{Dinamicresponsfunc&heligap}, resulting in the observation of re-entrant conductance feature \cite{heligapExper}. 

Also, spin polarized Luttinger liquids have also been investigated \cite{spinpolarizedLL}. The spin gap can be enhanced by strong electron-electron interactions due to the combined effect of Rashba SO and Zeeman magnetic field \cite{6Aseev2017intro0,6Aseev2017intro} or, equivalently, a spiral magnetic field \cite{spiralLL1,spiralLL2}, resulting in a perfect spin filter. The realization of spin filtering considering the backscattering term has been studied in the finite-size spinful wire connected to noninteracting leads \cite{10visooriGiama2020}. However, at low densities, there exists a critical field, depending on the electron density, above which the spin gap can be established \cite{heligapatlow_e_density}. It has been demonstrated that for interacting SO coupled quantum wires the spin-density-wave (SDW) state can be revealed in the presence \cite{eeSOC1DwithMag1,eeSOC1DwithMag2} or absence \cite{eeSOC1DwithoutMag} of an applied magnetic field or due to a modulated SO coupling \cite{eeModuSOC1D1,eeModuSOC1D2}. Moreover, strongly anisotropic electron-spin susceptibility \cite{14spinsusceptibilityhelical_Meng-Loss} and fractional conductance \cite{5aseev2018} have been revealed in nanowires due to the interplay between the Zeeman field and the SO term \cite{eeSOC1DwithMag31,eeSOC1DwithMag32,eeSOC1DwithMag33}.

At the edge of quantum spin Hall systems \cite{2Sooriintro,2Sooriintro1}, the effect of electron-electron interactions \cite{eeSpinHall} in the presence of a point contact \cite{eePoinyCont} or nonuniform Rashba SO \cite{eeNonUni} has been studied. Furthermore, nuclear-spin-induced edge resistance due to electron-electron backscattering has been predicted \cite{eenuclear-spin1,eenuclear-spin2}. The breaking of time-reversal symmetry in these systems due to the application of a magnetic field causes strong anisotropic feedback \cite{HelicalZeemanHam1,HelicalZeemanHam2}. Such a situation can be created either intrinsically by nuclear spins \cite{4aseev2017,11Hsu-nuclear-chargetransp} or extrinsically via the proximity effect through coupling to ferromagnetic materials \cite{FeroHetro,TM-REHetro1,TM-REHetro2}. These phenomena lead to a gap in the edge states \cite{2Sooriintro}, mixing the helicity of the states \cite{HelicalZeemanHam1}. Thus, the question of how electron-electron interactions affect various correlations and conductivities in 1D partially mixed helical states remains unanswered.

In this work, we investigate the low-energy properties of single-mode helical states with electron-electron interactions under the influence of an applied Zeeman exchange field, resulting in the opening of a gap in the spectrum. Notably, unlike in SO coupled nanowires, the gap in this scenario is global and arises from the partial mixing of helical spin states near the band edge \cite{HelicalZeemanHam1}. We employ a RG analysis to derive flow equations, enabling us to delineate the phase diagram and refine the bare transport calculations. Our analysis reveals three distinct phases: strong gap, crossover, and weak gap regimes, with the gap term becoming relevant in the strongly interacting regime. By considering the helicity in the x-direction, the Zeeman-generated gap in the y-direction, we investigate the charge-density-wave (CDW) and SDW correlation functions in both the strong and weak gap regimes, quantifying the corrections introduced by the gap term to these correlations. In the strong gap scenario, we observe substantial corrections in the components of the SDW along both the Zeeman field and helicity axes, although the corrections along the former axis dominate over those along the latter. Conversely, in the weak gap scenario, we distinguish between the exponential (y, z, yz, zy components of SDW) and non-exponential (CDW and x component of SDW) parts in the unperturbed correlation. For the exponential terms, we incorporate logarithmic corrections, while for the non-exponential components, we employ direct perturbative expansion methods.

Additionally, we calculate the electron charge conductivity in the mixed helical interacting system. Our approach involves employing a perturbative treatment known as the Memory function method. We enhance the accuracy of our conductivity calculations by incorporating the RG analysis, which is applicable across both low and high-temperature regimes. In the above gap, the results show a power-low conductivity improved by RG. However, at temperatures below the gap, an exponential behavior dependent on various parameters such as interactions and Fermi level becomes apparent. Specifically, when the Fermi level is higher than the gap, the conductivity exhibits an upward trend. Conversely, as the Fermi level lies below the gap, strongly repulsive interactions lead to the dominance of the strong gap phase, resulting in an exponential decay with respect to the normalized gap-temperature ratio.

The paper is organized as follows. In Sec. \ref{s2}, the system model and its Hamiltonian are presented. Using renormalization group analysis, flow equations of the system parameters as well as the phase diagram are obtained in Sec. \ref{s3}. In Sec. \ref{s4}, the correlation functions of charge and spin density waves are investigated in the relevant and irrelevant gap regimes. In the irrelevant gap regime, two complementary methods, logarithmic and residual corrections are employed for exponential and non-exponential operators, respectively. Using Memory function method, charge conductivity is studied in Sec. \ref{s5}. Also, concluding remarks are summarized in Sec. \ref{s6}. Some details of derivations are placed in the Appendices.

\section {Model}\label{s2}

\begin{figure}[t!]
    \centering
    \includegraphics[width=8cm]{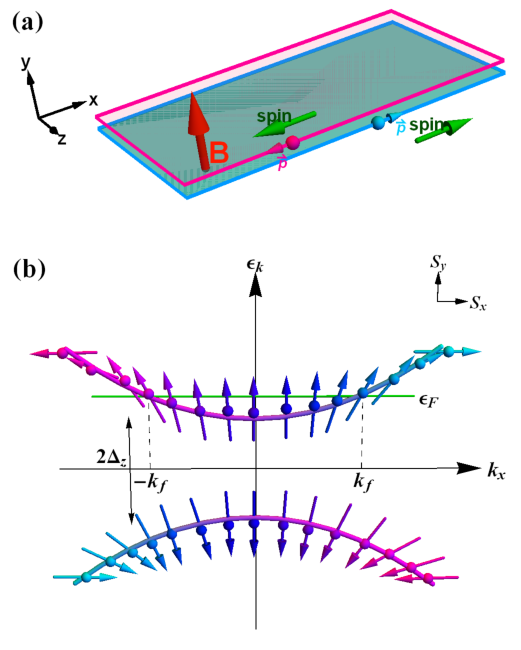}
   \caption{(Color online) (a) Setup of two counter-propagating helical modes in the presence of the Zeeman field being perpendicular to the surface of the sample. (b) The band structure of the gapped helical states (with gap 2$\Delta_z$) along with its spin states showing a half-spiral pattern in each band due to mixing of the right-going spin states (turquoise color) with the left-going ones (pink color) at low energies.}
    \label{fig1}
\end{figure}

We consider a 1D helical state at the edge of a 2D TI along the x direction subjected to a perpendicular Zeeman exchange field as shown in Fig. \ref{fig1}(a). The Zeeman field is so weak that cannot wash out the edge states and only can mix the edge states by gapping them out, see Fig. \ref{fig1}(b). Using the effective edge model \cite{Ts1}, the low-energy Hamiltonian of the 1D gapped helical states is \cite{1DGapFormImpu,3Soori2012,HelicalZeemanHam1,HelicalZeemanHam2,GappedHelicalHam}
\begin{equation}
\mathcal{H}_0=\hbar v_F\psi^{\prime\dagger}(x)k_x\sigma_x\psi^\prime(x)+\Delta_z\psi^{\prime\dagger}(x)\sigma_y\psi^\prime(x),
\label{Ham0}
\end{equation}
where $v_F$ is the Fermi velocity, $k_x$ is the wave vector along the edge, $\sigma_{x,y}$ are the Pauli matrices in spin space and $\Delta_z=g\mu_BB_z/2$ is the Zeeman gap with $g,\mu_B$, and $B_z$ denoting g-factor, Bohr magneton, and magnetic field in the z direction respectively. Also, we take $\hbar=1$. Here, we have defined the single-electron field operator
\begin{equation}
\psi^\prime(x)=\sum_r\chi_r(rk_F)\psi_r(x)e^{irk_Fx},
\label{fieldoperator}
\end{equation}
with $\psi_r^{(\dagger)}(x)$ is the annihilation (creation) field operators of the right-going ($r=R$) and left-going ($r=L$) carriers at the position $x$ with the Fermi momenta $\pm{k_F}=\pm\sqrt{\epsilon_F^2-\Delta_z^2}/v_F$ and the Fermi energy $\epsilon_F$. Also, the spinor function is
\begin{equation}
\chi_r(k)=\frac{1}{\sqrt{2}}\begin{pmatrix}
re^{i\vartheta_k}\\e^{-i\vartheta_k}
\end{pmatrix},
\end{equation}
with
\begin{equation}
\vartheta_k=\frac{1}{2}\arctan(-\frac{\Delta_z}{kv_F}).
\end{equation}
Note that $\vartheta_k$ would mix the different spin states around the bands' edge if $\Delta_z \neq 0$, as shown in Fig. \ref{fig1}(b). 

The Hamiltonian of electron-electron interactions is 
\begin{align}
\mathcal{H}_{int}=&g_1\psi_L^{\prime\dagger}(x)\psi_R^{\prime\dagger}(x)\psi_L^\prime(x)\psi_R^\prime(x)\nonumber\\
+&g_2\psi_L^{\prime\dagger}(x)\psi_L^\prime(x)\psi_R^{\prime\dagger}(x)\psi_R^\prime(x)\nonumber\\
+&\frac{g_4}{2}\sum_{r=R,L}[\psi_r^{\prime\dagger}(x)\psi_r^\prime(x)]^2,
\end{align}
where $g_1$ is the backscattering interaction constant. Also, $g_2$ and $g_4$ are the dispersive and forward interaction constants, respectively.

Typically, the application of bosonization requires certain criteria to be met, including the system's one-dimensionality, the absence of long-range order, the presence of power-law correlations, and being in the low-energy and continuum limits. In our study, our system fulfills all these conditions, rendering the bosonization representation applicable for our analysis. 

Using the standard method of the bosonization, expressing the fermionic operators in terms of bosonic fields with $\vartheta_k=0$, we arrive at the total Hamiltonian $\mathcal{H}=\mathcal{H}_0+\mathcal{H}_{int}$ as
\begin{equation}
\begin{split}
\mathcal{H}=\frac{v}{2}[\frac{1}{K}(\partial_x\Phi)^2+K(\partial_x\Theta)^2]+\frac{\lvert{v_F k_F}\rvert}{\sqrt{4\pi}}\partial_x\Phi\\+\frac{\Delta_z}{\pi a_0}\cos(\sqrt{4\pi}\Phi+2k_Fx),
\label{Hamfirst}
\end{split}
\end{equation}
where $\Phi =\phi_R+\phi_L$ and $\Theta =\phi_R-\phi_L$ 
with $\phi_R,\phi_L$ being the chiral bosonic fields. $a_0$ is an infinitesimal ultraviolet cutoff. Also, the Luttinger liquid parameter $K$ and the velocity of collective mode $v$ are found as
\begin{equation}
K=\sqrt{\frac{v_F -\frac{g_{fb}}{8\pi}+\frac{g_4}{8\pi}}{v_F +\frac{g_{fb}}{8\pi}+\frac{g_4}{8\pi}}},
\label{Luttingerparameter}
\end{equation}
\begin{equation}
v=\sqrt{(v_F +\frac{g_4}{8\pi})^2-(\frac{g_{fb}}{8\pi})^2}.
\label{Luttingervelocity}
\end{equation}
Here, we have defined $g_{fb}=g_2-g_1$. Note that $\Delta_z$, mixing the helicity, provides the sine-Gordon term in the Hamiltonian.

To rewrite Hamiltonian (\ref{Hamfirst}) in the conventional form, we absorb the gradient term into the kinetic part of the Hamiltonian by shifting the field $\Phi$ as
\begin{equation}
\tilde{\Phi}=\Phi+\frac{\lvert{v_F k_F}\rvert}{\sqrt{4\pi}}\frac{K}{v}x. 
\label{tildePhi}
\end{equation}
This transformation also changes the cosine term, so we rewrite it as
\begin{equation}
\mathcal{H}_{\Delta_z}=\frac{\tilde{\Delta}_z}{2\pi a_0}e^{i\sqrt{4\pi}\tilde{\Phi}}+h.c.,
\label{SGor}
\end{equation}
where $\Delta_z$ is replaced with $\tilde{\Delta}_z$;
\begin{equation}
\tilde{\Delta}_z=\Delta_z e^{-i\alpha x},
\end{equation}
with
\begin{equation}
\alpha=\lvert k_F\rvert K^2-2k_F,
\label{KAlpah}
\end{equation}
where we have used $vK=v_F$. Therefore, the total Hamiltonian $\mathcal{H}=\mathcal{H}_{hel}+\mathcal{H}_{\Delta_z}$ reads
\begin{equation}
\mathcal{H}=\frac{v}{2}[\frac{1}{K}(\partial_x\tilde{\Phi})^2+K(\partial_x \Theta)^2]+\frac{\tilde{\Delta}_z}{2\pi a_0}e^{i\sqrt{4\pi}\tilde{\Phi}}+h.c.,
\label{Hamsecond}
\end{equation}
where $\mathcal{H}_{hel}$ represents the Hamiltonian of the helical Tomonaga-Luttinger liquid.

\section{Renormalization Group analysis}\label{s3}

To understand the role of perturbations on the low-energy properties of our 1D gapped Dirac system in the framework of field theory, the RG analysis can be employed. To do so, we obtain the helical Tomonaga-Luttiger liquid action via integrating out the $\Theta$ field in the quadratic part of Eq. (\ref{Hamsecond}), $\mathcal{H}_{hel}$,
\begin{equation}
S_{hel}=\frac{1}{2K}\int d^2r [\frac{1}{v}(\partial_\tau\tilde{\Phi})^2+v(\partial_x\tilde{\Phi})^2],
\end{equation}
with the bosonic field $\tilde\Phi{(r,\tau)}$ in the imaginary time $\tau$. Correspondingly, the contribution to the imaginary-time action from $\mathcal{H}_{\Delta_z}$ is
\begin{equation}
S_{\Delta_z}=\frac{\Delta_z}{2\pi a_0}\int d^2r e^{-i\alpha x}e^{i(\sqrt{4\pi}\tilde{\Phi})}+h.c..
\end{equation}
So, one can get the effective action $S_{eff}=S_{hel}+S_{\Delta_z}$. Following standard methods \cite{Giamarchi}, we use the effective action $S_{eff}$ to compute the correlation function,
\begin{equation}
\begin{split}
\langle e^{i\sqrt{4\pi}[\Phi(r_1)-\Phi(r_2)]}\rangle_{S_{hel}+S_{\Delta_z}}\equiv \mathcal{Z}^{-1}\\\times\int D\Phi e^{-S_{hel}}e^{-S_{\Delta_z}}e^{i\sqrt{4\pi}[\Phi(r_1)-\Phi(r_2)]},
\end{split}
\end{equation}
where $\mathcal{Z}\equiv \int D\Phi e^{-(S_{hel}+S_{\Delta_z})}$ is the partition function. We expand the correlation function up to the second-order terms in $\Delta_z$. The non-perturbed correlation function reads as
\begin{equation}
    \langle e^{i\sqrt{4\pi}[\Phi(r_1)-\Phi(r_2)]}\rangle_{S_{hel}}=e^{-i(\alpha+2k_F)(x_1-x_2)-2KF_1(r_1-r_2)},
\end{equation}
where 
\begin{equation}
\begin{split}
    F_1(r_1-r_2)=\frac{1}{2}log[\frac{(x_1-x_2)^2+(v(\tau_1-\tau_2))^2}{a_0^2}]\\+\frac{t}{K}\cos(2\theta_{r_1-r_2}),
    \label{F1}
\end{split}
\end{equation}
with $\theta_r$ is the angle between the space-time vector $(r,v\tau)$ and the space coordinate axis $r$, and $t$ indicates the anisotropy strength between the space and time directions. Furthermore, we find that the perturbtive correlation function has the same form as the free one with the effective quantities $e^{-i(\alpha^{eff}+2k_F)(x_1-x_2)-2K^{eff}F_1^{eff}(r_1-r_2)}$, where $F_1^{eff}(r_1-r_2)$ is the same as Eq. (\ref{F1}) but with the following effective parameters
\begin{widetext}
    \begin{align}
K^{eff}&=K-\pi^2\frac{\mathcal{Y}^2K}{2log[\frac{\lvert r_1-r_2\rvert^2}{a_0^2}]}2(r_1-r_2)\int_{a_0}^{\infty}\frac{dr}{a_0^2}(\frac{r}{a_0})^{2-2K}J_1(\alpha r)-\pi^2\frac{\mathcal{Y}^2K^2}{4}\int_{a_0}^{\infty}\frac{dr}{a_0^2}(\frac{r}{a_0})^{3-2K}J_0(\alpha r),\\
t^{eff}&=t+\pi^2\frac{\mathcal{Y}^2K^2}{8}\int_{a_0}^{\infty}\frac{dr}{a_0^2}(\frac{r}{a_0})^{3-2K}J_2(\alpha r),\\
\alpha^{eff}&=\alpha+\pi\frac{\mathcal{Y}^2K}{2(x_1-x_2)}2(r_1-r_2)\int_{a_0}^{\infty}\frac{dr}{a_0^2}(\frac{r}{a_0})^{2-2K}J_1(\alpha r),
\end{align}
\end{widetext}
where $\mathcal{Y}=\frac{2\sqrt{2}\Delta_z a_0}{\pi v}$ is the dimensionless gap constant and $J_{n}(x)$ is the Bessel function of order $n$. 

Upon transitioning from the original cutoff $a_0$ to the parameterized cutoff $a_0(l)=a_0e^l$, where $l$ denotes the length scale and undergoes a shift from $l$ to $l+dl$, we derive the following set of RG flow equations,
\begin{align}
\frac{d\mathcal{Y}(l)}{dl}&=(2-K(l))\mathcal{Y}(l),
\label{flowgap}
\\
\frac{d\alpha(l)}{dl}&= \frac{\mathcal{Y}^2(l) K(l)}{2\pi}\alpha(l),
\label{flowalpha}\\
\frac{dK(l)}{dl}&=-\frac{1}{4}\mathcal{Y}^2(l) K^2(l)J_0(\alpha(l) a_0),
\label{KRG}\\
\frac{dt(l)}{dl}&=\frac{\mathcal{Y}^2(l)}{8}K(l)^2J_2(\alpha(l)a_0).
\label{tRG}
\end{align}
Notably, from Eq. (\ref{flowgap}), we observe that the gap associated with the sine-Gordon term exhibits RG relevance when $K<2$. According to Eqs. (\ref{flowgap}) and (\ref{flowalpha}), the evolution of $\mathcal{Y}(l)$ remains unaffected by $\alpha(l)$, despite $\mathcal{Y}(l)$ influencing the renormalization of $\alpha(l)$. Similarly, from Eqs. (\ref{flowalpha}) and (\ref{KRG}), one can see that $K(l)$ impacts the renormalization of $\alpha(l)$, while the flow of $K(l)$ remains unaltered by $\alpha(l)$ except for its dependence on $J_0(\alpha a_0)$.

\begin{figure}[t!]
    \centering
    \includegraphics[width=8.5 cm]{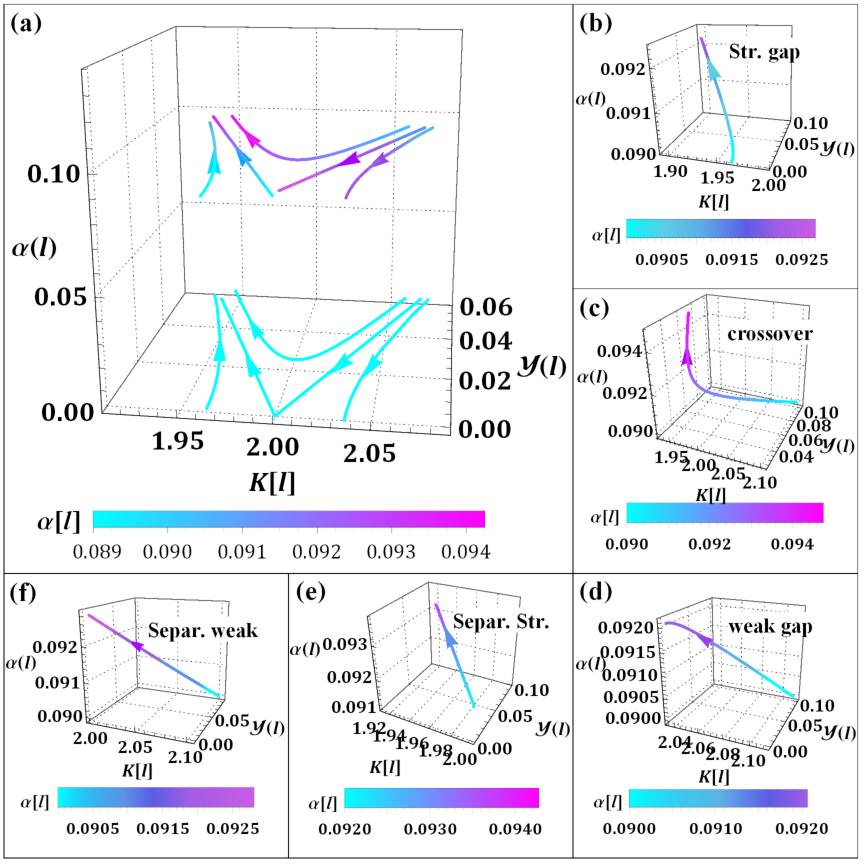}
   \caption{(Color online) (a) RG flows of the gapped helical model in the presence of the gap-deviation parameter. For the parameters in the below separatrix (straight line with purple color) with $K>2$, the gap coefficient vertically flows to zero (pink color). for $K<2$, in the above separatrix, $\mathcal{Y}$ flows to the strong coupling. The turquoise lines are projects of the flow trajectories with $\alpha(l)=0$. Individual flow in (b) strong, (c) crossover, and (d) weak gap regimes. Individual path of the separatrix for (e) $K<2$ and (f) $K>2$.}
    \label{fig2}
\end{figure}

Numerical solutions to the RG equations (\ref{flowgap})-(\ref{KRG}) are shown in Fig. \ref{fig2}. From the $(K(l), \mathcal{Y}(l))$ plane of Fig. \ref{fig2}(a), one finds three regimes separated by the separatrix (the straight lines) with a fixed point at $(K, \mathcal{Y})=(2,0)$: (i) strong gap, (ii) crossover, and (iii) weak gap regimes. These regimes are respectively located below the separatrix with $K<2$, above the separatrix lines, and below the separatrix with $K>2$. In the strong gap regime, $\mathcal{Y}(l)$ tends to infinity with increasing length scale $l$. In the crossover regime, $\mathcal{Y}(l)$ initially decreases and then tends towards the strong gap regime. In the weak gap regime, the coupling $\mathcal{Y}(l)$ approaches zero as $l$ tends to infinity, rendering the gap term in the Hamiltonian irrelevant. It is notable that in the strong gap and crossover regimes, the gap term is relevant in the renormalization process. Consequently, strong repulsive interactions can mix the spin states due to promoting the gap. 

On the other hand, $\alpha$ renormalizes weakly. For the sake of clarity, a close view, exhibiting the flow of $\alpha(l)$, is illustrated in Figs. \ref{fig2}(b)-\ref{fig2}(d), in the strong gap, crossover, and weak gap regimes, respectively. Furthermore, the $\alpha$ dependence of the separatrix lines for $K<2$ and $K>2$ is shown in Figs. \ref{fig2}(e) and \ref{fig2}(f), respectively. These figures illustrate that $\alpha$ renormalizes towards larger values as $K$ goes from the weakly interacting regime to the strongly interacting one. This behavior can be comprehended from the flow equation (\ref{flowalpha}). It suggests that whether in the weak gap or strong gap regime, the coefficient of the parameter $\alpha$ on the right side of the equation is consistently positive. Consequently, $\alpha$ will invariably increase as the length scale $l$ increases, irrespective of the gap values. This explains the upward trend observed in each of the flow paths across different regimes. 

Note that due to Eq. (\ref{F1}), the flow of the space and time anisotropy parameter $t$, as described by Eq. (\ref{tRG}), results in the renormalization of the velocity of the excitations, given by $\frac{dv(l)}{dl}=-\frac{\mathcal{Y}^2(l)}{4}v(l)K(l)J_2(\alpha(l)a_0)$. The magnetic field in the relevant gap regime plays a role in decreasing the velocity of the boson excitations. On the other hand, the impact of interactions can be observed through the Luttinger parameter $K$ in the velocity's flow equation. This implies that strong interactions, in conjunction with a relevant gap, can influence the velocity. Therefore, the magnetic field can induce a gap that is further modified by the interactions. This phenomenon bears resemblance to the umklapp process, where interactions are responsible for creating a gap in the system, consequently reducing the Fermi velocity due to the commensurate potential \cite{Giamarchi}. 

Remarkably, unlike in previous studies \cite{umklapp-Giamarchi,spiralLL1,spiralLL2,10visooriGiama2020,Citro-Giamarchi}, in the present case, the sine-Gordon model does not arise from electron-electron interactions. Consequently, electron-electron interactions are not directly responsible for creating the gap. However, they do play a role in influencing the various gap regimes. To investigate the effect of interactions on the relevance of the gap, we introduce,
\begin{equation}
K=2-8\mathcal{Y}_{int},
\label{yint}
\end{equation}
where $\mathcal{Y}_{int}=\frac{g_{fb}+1}{8}$ represents the gap-tuning parameter of the interaction. Here, we have considered $g_{fb}=g_4$. When $g_{fb}\rightarrow-1$, we find the limit of the marginal regime. In the absence the interaction, $g_{fb}=0$, the Luttinger parameter is $K=1$, indicating a relevant regime. Furthermore, in the presence of interaction within the attractive regime $-1<g_{fb}<0$, we observe that $1<K<2$, thus indicating the persistence of the relevant regime. However, if $g_{fb}<-1$, then $K>2$, leading to the irrelevant regime where $\mathcal{Y}$ tends to zero. At $K=2$, a change in the flow occurs, indicating the marginal regime. When the interaction becomes repulsive, $g_{fb}>0$, then $K<1$, placing us deeply in the relevant regime, where the stronger repulsive interaction enforces the gap to be more relevant.

\section{Charge and spin density
correlation functions}
\label{s4}

In this section, we explore charge and spin density correlation functions in the presence of the gap and electron-electron interaction. In a Luttinger liquid, an interacting one-dimensional system, low-energy excitations are not described by quasi-particles but rather by collective density waves that carry charge and spin degrees of freedom independently. This leads to the formation of charge and spin density wave phases. We investigate such correlations in a helical wire where the separation of spin and charge is slightly broken, leading to spin-momentum coupling. We then compare this with the helical wire under a Zeeman field, which creates a slight mixing in the helical edges at $k=0$, disturbing the momentum-spin coupling. We aim to understand how the charge and spin density of the system respond to this disturbance and how interactions influence the conditions of the problem. Specifically, we seek to determine how these collective excitations modify the power-law behavior of the Luttinger liquid at the Fermi level, identifying which excitations become dominant and which are suppressed. To do so, we consider the CDW and different components of the SDW operators, respectively, given by
\begin{align}
\mathcal{O}_{CDW}(x) &= \psi^{\prime \dagger}(x) \psi^{\prime}(x), \\
\mathcal{O}^{i}_{SDW}(x) &= \psi^{\prime \dagger}(x) \sigma_{i} \psi^{\prime}(x), \quad i = x, y, z
\end{align}
where $\psi^{\prime}(x)$ is the single-electron field operator defined in Eq. (\ref{fieldoperator}) and $\sigma_{i}$ are Pauli matrices. Using standard bosonization techniques, the operators can be written in terms of the bosonic fields, yielding
\begin{align}
 \mathcal{O}_{CDW}(x,\tau)&=-\frac{1}{\sqrt{4\pi}}\partial_x\Phi(x,\tau),
 \label{CDWOp}
 \\\mathcal{O}^{x}_{SDW}(x,\tau)&=-\frac{1}{\sqrt{4\pi}}\partial_x\Theta(x,\tau),
  \label{xSDWOp}
  \\
 \mathcal{O}^{y}_{SDW}(x,\tau)&=\frac{1}{\pi a_{0}}\cos(\sqrt{4\pi}\Phi(x,\tau)+2k_{f}x),
   \label{ySDWOp}\\
 \mathcal{O}^{z}_{SDW}(x,\tau)&=-\frac{1}{\pi a_{0}}\sin(\sqrt{4\pi}\Phi(x,\tau)+2k_{f}x).
 \label{zSDWOp}
\end{align}

Furthermore, the charge and spin correlations are defined as \cite{Giamarchi},
\begin{align}
\mathcal{R}_{CDW}(r_1,r_2)&=\langle T_{\tau}\mathcal{O}_{CDW}(r_1)\mathcal{O}_{CDW}(r_2)\rangle,\\
\mathcal{R}^{ij}_{SDW}(r_1,r_2)&=\langle T_{\tau}\mathcal{O}^i_{SDW}(r_1)\mathcal{O}^j_{SDW}(r_2)\rangle,
\end{align}
where $T_{\tau}$ is the time ordering operator, $r_1=(x_1,v\tau_1)$, and $r_2=(x_2,v\tau_2)$. Without loss of generality, here after, we take $r_1=r$ and $r_2=0$. Plugging Eqs. (\ref{CDWOp})-(\ref{zSDWOp}) into the above definitions and using Eq. (\ref{tildePhi}), the non-perturbative charge and spin correlations with respect to $\mathcal{H}_{hel}$ can be obtained as
\begin{align}
\mathcal{R}_{CDW}(r)&=-\frac{1}{4\pi^2}KF_2(r),
\label{CDWnonperturb}\\
\mathcal{R}^{xx}_{SDW}(r)&=-\frac{1}{4\pi^{2}K}F_2(r),\label{xSDWnonperturb}\\
\mathcal{R}^{yy}_{SDW}(r)&=\frac{\cos(\alpha x)}{2(\pi a_0)^2}e^{-2KF_{3}(r)},\label{ySDWnonperturb}\\
\mathcal{R}^{zz}_{SDW}(r)&=\frac{\cos(\alpha x)}{2(\pi a_0)^2}e^{-2KF_{3}(r)},\label{zSDWnonperturb}\\
\mathcal{R}^{yz}_{SDW}(r)&=-\mathcal{R}^{zy}_{SDW}(r)=-\frac{\sin(\alpha x)}{2(\pi a_0)^2}e^{-2KF_{3}(r)},
\label{yzSDWnonperturb}
\end{align}
where 
\begin{align}
F_2(r)&=\frac{(v\tau sign(\tau)+a_0)^{2}-x^2}{2((v\tau sign(\tau)+a_0)^{2}+x^2)^2},
\label{F2}\\
F_{3}(r)&=\frac{1}{2}log(\frac{(v\tau sign(\tau)+a_0)^{2}+x^2}{a_0^{2}}).
\label{F3}
\end{align}
Note that the mixed correlations $\mathcal{R}^{xy}_{SDW}(r)$ and $\mathcal{R}^{xz}_{SDW}(r)$ have no contribution. Because they are of the form $\langle \partial_x\Theta(x)e^{\sqrt{4\pi}\Phi(x)+2k_{f}x}\rangle$ being zero.

In the undisturbed state of the helical Luttinger liquid, both the charge and the x component of correlations exhibit similar behavior. This similarity arises from the coupling between the charge's momentum and the x component of spin within the helical regime. However, a different scenario reveals for the y and z components, as well as the yz and zy components, where an interaction-dependent power-law behavior is observed. This behavior originates from the presence of helicity in the x direction, which permits the mixing of up and down orientations of spins in each of the chiral modes for the y and z directions. Therefore, in these two directions, we observe a normal Luttinger liquid behavior, albeit with the additional manifestation of interaction traces in the coefficient of the correlation functions, alongside their non-universal power-law scaling.

Since the gap has both strong and weak regimes, the density wave correlations in the presence of gap can be evaluated in the different gap regimes. First, we find corrections to the spin and charge correlation functions in the strong gap regime and then we calculate them in the weak gap case. In the weak gap limit, we decompose the non-perturbative correlation functions into exponential and non-exponential parts. For the exponential terms, we apply the logarithmic correction \cite{Giamarchi,Giamarchi-schulz}, while for the others, we employ the perturbative expansion method directly \cite{Citro-Giamarchi}. 

\subsection{Corrections in the strong gap regime}

In the relevant regime, the field $\tilde\Phi$ orders. The order of field $\tilde\Phi$ corresponds to the order of the field $\Phi$. Therefore the dual field $\Theta$ is totally disordered and all its long-range components, if any, will be zero. Knowing that the characteristic parameter of the gap, $\Delta_z$, is positive, our perturbation Hamiltonian minimizes when the gapped field $\tilde\Phi$ is,
\begin{equation}
\tilde\Phi=\sqrt{\pi}(n+\frac{1}{2}),
\end{equation}
where $n$ is an integer. So, since $\langle\cos(\sqrt{4\pi}\tilde\Phi-\alpha x)\rangle\neq0$, the y component of the SDW (\ref{ySDWOp}) will have a power-law behavior and develop the long-range order in the system. While, because of $\langle\sin(\sqrt{4\pi}\tilde\Phi-\alpha x)\rangle=0$, the z component of the SDW (\ref{zSDWOp}) becomes zero. This means that there is no backscattering associated with spin flipping in the z direction. Also, the CDW (\ref{CDWOp}) becomes zero. The x component of the SDW (\ref{xSDWOp}), which includes the field gradient $\Theta$, continues to behave as Fermi liquid like, but with strong gap values for the Luttinger parameter. This implies that the helical order in the x direction is weakly disrupted by the gap.
    
As already discussed, the gap induced by the Zeeman field is RG relevant when $K<2$, leading to the gapping out of the field $\tilde\Phi(r)$ and perturbing the helical spin order. In the relevant sense, the RG flow is integrated until the length scale $\frac{v(l)}{\Delta_z(l)}$ associated with the running gap $\Delta_z(l)$ of $\tilde\Phi(r)$ equals the running short-distance cutoff. Namely, $\Delta_z(l)$ is the greatest the energy scale;
    \begin{equation}
        e^{l^*}=\frac{v(l^*)/\Delta_z(l^*)}{a_0(l^*)}\sim{1}.
    \end{equation}
This boundary is the end of the validity of the flow. In this limit, one can expand the sine-Gordon term of Hamiltonian (\ref{Hamsecond}) to the second order yielding 
\begin{align}
H(k)=&\frac{v^*}{2}[\frac{1}{K^*}(\partial_{x}\tilde\Phi)^2+K^*(\partial_{x}\Theta)^2]+\frac{2\Delta_z}{a_0}\tilde{\Phi}^2\nonumber\\
&+\frac{2\Delta_z}{\sqrt{\pi}a_0}(2k_{f}x-v_f k_f\frac{K^*}{v^*}x)\tilde{\Phi},
\end{align}
where $v^*$ and $K^*$ are the strong coupling values of the velocity and the Luttinger liquid parameters as $v(l^*)=\Delta_z(l^*){a_0}^*$. With the action constructed by this Hamiltonian and the formation of a Gaussian partition function, according to Ref. \cite{Giamarchi}, we obtain the basic correlation as
\begin{equation}
\langle\tilde{\Phi}^{\dagger}(q_1)\tilde{\Phi}(q_2)\rangle=\frac{\pi K^*\Omega\beta}{\frac{\omega_{n1}^2}{v^*}+v^*k_1^2+\frac{4\pi K^*\Delta_z}{a_0}}\delta_{q_1,q_2},
\end{equation}
where $\Omega$ is the volume of system, $q_i=(k_i,\omega_{ni}/{v^*})$ is the Fourier coordinate of space-time and $\tilde{\Phi}(q_i)$ is the Fourier transform of the bosonic field $\tilde\Phi$. Also, $\beta=k_B T$ with $k_B$ and $T$ being Boltzmann constant and temperature, respectively. For simplicity, we set $k_B=1$. Therefore, the effective Hamiltonian plays the same role as the quadratic Hamiltonian during the averaging for the correlations.
Now, one can calculate corrections to the correlation functions using Eqs. (\ref{CDWOp})-(\ref{zSDWOp}) and the Harmonic perturbative action. The leading-order corrections in the limit $r\sim{a_0}$ and $\Delta_z \gg 1$ can be obtained as
\begin{align}
\mathcal{R}^{str,xx}_{SDW}(r)&=\frac{1}{4\pi^2 K^*}F^*_2(r),\\
\mathcal{R}^{str,yy}_{SDW}(r)&=\frac{\cos(\alpha a_0)}{2(\pi a_0)^2},
\end{align}
where $F^*_2(r)=\frac{(v^*\tau sign(\tau)+a_0)^{2}-x^2}{2((v^*\tau sign(\tau)+a_0)^{2}+x^2)^2}$. The x component correction of the SDW persists in its helical existence but is influenced by the presence of the gap. The constant value for the y component means that this phase remains stable in the strong gap regime. Moreover, regularization of the $\tilde{\Phi}$ field halts its temporal and spatial variations, resulting in the disappearance of the CDW's correction. 
The corrections of z component and its mixture suppress exponentially. The reason is that the field $\tilde{\Phi}$ orders, thus the operators including  $\sin(\sqrt{4\pi}\tilde{\Phi}(r)-\alpha x)$ vanish identically. As a result, when the gap becomes relevant, the mixed helical states lose their capability of coupling charge and spin, leading to the dominance of the y-component of the SDW phase, which stabilizes towards a constant value.

\subsection{Corrections in the weak gap regime}
\label{IV-2}
\subsubsection{Logarithmic corrections}
\label{IV-2-1}

When the operators are marginal, the logarithmic corrections to the correlation functions should be taken into account \cite{Giamarchi}. In this procedure, we write the renormalization equations for the correlation functions themselves. Using Eq. (\ref{yint}), we rewrite RG Eqs. (\ref{flowgap})-(\ref{tRG}) in terms of $\mathcal{Y}_{int}$ as
\begin{align}
\frac{d\mathcal{Y}(l)}{dl}&=8\mathcal{Y}_{int}(l)\mathcal{Y}(l),
\label{yRG}\\
\frac{d\alpha(l)}{dl}&=\frac{\mathcal{Y}^2(l) }{\pi}\alpha(l),
\label{alphaRG}\\\frac{d\mathcal{Y}_{int}(l)}{dl}&=\frac{1}{8}\mathcal{Y}^2(l),
\label{yintRG}
\\
\frac{dt(l)}{dl}&=\frac{\mathcal{Y}^2(l)}{2}J_2(\alpha(l)a_0).
\label{tyRG}
\end{align}
These equations correspond to an expansion up to second order in the interaction parameter $\mathcal{Y}_{int}$. From (\ref{yintRG}) and (\ref{yRG}), we obtain $\mathcal{Y}\frac{d\mathcal{Y}}{dl}=64\mathcal{Y}_{int}\frac{d\mathcal{Y}_{int}}{dl}$ and thus we get a constant of motion as
\begin{equation}
    A^2=64\mathcal{Y}_{int}^2-\mathcal{Y}^2.
    \label{constmotion1}
\end{equation}
Also, using Eqs. (\ref{yintRG}), (\ref{yRG}), and (\ref{alphaRG}), we obtain the second constant of motion
\begin{equation}
    B^2=\frac{\mathcal{Y}(l)^2}{2}-\frac{\pi^2}{2}log[\alpha(l)]^2.
    \label{constmotion2}
\end{equation}
The two constants of motion (\ref{constmotion1}) and (\ref{constmotion2}) determine the regions with different regimes associated with their boundaries. Inserting the constants of motion in Eqs. (\ref{yRG}), (\ref{alphaRG}) and (\ref{yintRG}) the flow can be integrated. For weak gap region with $8\mathcal{Y}_{int}>\mathcal{Y}$ and $\mathcal{Y}>\pi log(\alpha)$ where $K>2$, one gets
\begin{align}
&\mathcal{Y}(l)=\frac{A}{\sinh[Al+\tanh^{-1}(\frac{A}{8\mathcal{Y}_{int}(0)})]},
\label{ysolution}\\
&\alpha(l)=\exp[\frac{\sqrt{2}B \tan[\sqrt{2}B l+\tan^{-1}(\frac{\pi}{\sqrt{2}B}log(\alpha(0)))]}{\pi}],
\label{alphasolution}\\&\mathcal{Y}_{int}(l)=-\frac{1}{8}\frac{A}{\tanh[Al+\tanh^{-1}(-\frac{A}{8\mathcal{Y}_{int}(0)})]}.
\label{yintsolution}
\end{align}

As already mentioned, when the gap turns on, the field $\tilde{\Phi}$ in the Hamiltonian (\ref{Hamsecond}) becomes ordered and the correlation functions containing this field decay exponentially, except for the CDW. In the sense of the logarithmic correction, we just consider Eqs. (\ref{ySDWnonperturb})-(\ref{yzSDWnonperturb}) exhibiting exponential behavior. The CDW correlation (\ref{CDWnonperturb}) and the x component of SDW correlation (\ref{xSDWnonperturb}) do not have any exponential term and so any logarithmic corrections. In the next section, we will calculate these correlations by applying second-order perturbation theory. Now, for instance, deriving the logarithmic correction of the y component of SDW correlation in perturbation yields (see Appendix \ref{logcorrect}),
\begin{align}
\mathcal{R}^{log,yy}_{SDW}(r)=\frac{\cos(\alpha(0) x)}{2(\pi a_0)^2}(\frac{a_0}{r})^{4+16\mathcal{Y}_{int}}e^{2t\cos(2\theta_r)}\nonumber\\\times L_1(r)L_2(r)L_3(r),
\label{finalLogCorr}
\end{align}
where
\begin{align}
L_1(r)&=\exp[32\int_{0}^{l_r}\mathcal{Y}_{int}(l) dl],\label{L1}\\
L_2(r)&=\exp[\frac{\cos(\alpha(l_r)x)}{\cos(\alpha(0)x)}-1],\label{L2}\\
L_3(r)&=\exp[-4\cos(2\theta_r)J_2(\alpha a_0)t(l_r)].
\end{align}
Here, $t(l_r)$ is the anisotropy parameter in the time-space and $L_3(r)$ is evaluated in Appendix \ref{logcorrect}. In what follows, we consider two general cases depending on whether the system is at the separatrix or not. 

In the first case, the system is considered at the line of the separation between the crossover and the weak regime, so that $\mathcal{Y}=8\mathcal{Y}_{int}$ and $\mathcal{Y}=\pi log(\alpha)$. The expression for the y component of the SDW can be found as
\begin{align}
\mathcal{R}^{log,yy}_{SDW}(r)=\frac{\cos(\alpha(0) x)}{2(\pi a_0)^2}(\frac{a_0}{r})^{4+16\mathcal{Y}_{int}}e^{2t\cos(2\theta_r)}\nonumber\\\times\mathcal{Y}_{int}^{4}log(\frac{a_0}{r})^{4}\exp[\frac{\cos(e^{\frac{-1}{\pi log(\frac{r}{a_0})}}x)}{\cos(\alpha(0)x)}-1].
\label{0logcorrectyy}
\end{align}
The z component is identical to the y component. The yz and zy components are similar to the $\mathcal{R}^{log,yy}_{SDW}(r)$ but with different coefficients,
\begin{align}
   \mathcal{R}^{log,zy}_{SDW}(r)=-\mathcal{R}^{log,yz}_{SDW}(r)=\frac{\sin(\alpha(0) x)}{2(\pi a_0)^2}(\frac{a_0}{r})^{4+16\mathcal{Y}_{int}}\nonumber\\\times e^{2t\cos(2\theta_r)} \mathcal{Y}_{int}^{4}\log(\frac{a_0}{r})^{4}\exp[\frac{\cos(e^{\frac{-1}{\pi log(\frac{r}{a_0})}}x)}{\cos(\alpha(0)x)}-1].
\label{0logcorrectyz}
\end{align}
The space-time dependence of the above equations indicates that the corrections of the correlation functions exhibit an oscillating damping behavior. Note also that the anisotropy parameter appears as an exponential factor. As $K$ increases from 2, $\mathcal{Y}_{int}$ becomes more negative, causing the power of $\frac{a_0}{r}$ to decrease. Consequently, the correlation functions persist even at larger distances. 

\begin{figure}[t!]
    \centering
\includegraphics[width=8.5cm]{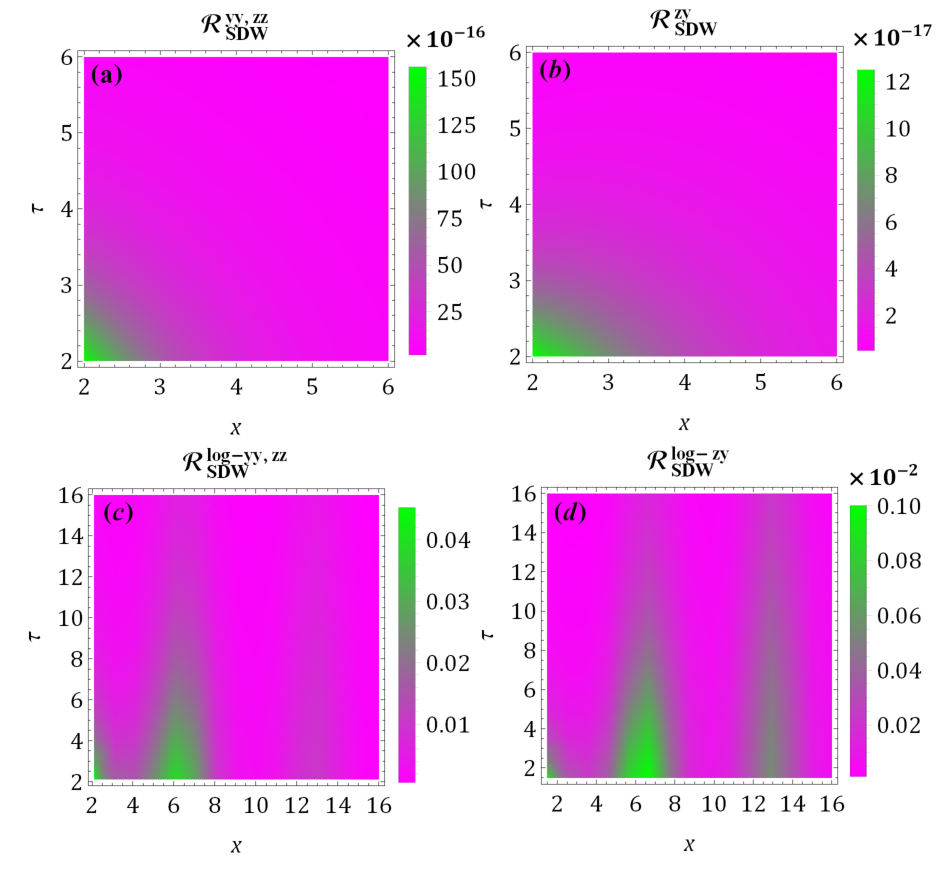}
   \caption{(Color online) Space-time dependence of the correlation functions schematic. The non-perturbative (a) y, z components of SDW and (b) zy ones. The logarithmic correction in the weak gap for (c) y, z components and (d) zy component of SDW.}
    \label{fig3}
\end{figure}

In Fig. \ref{fig3}, the space-time dependence of the SDW correlation functions is depicted. The top panels show the unperturbed components of the SDW correlations. One can observe a space-time isotropy, with the SDWs decreasing as $x$ and $\tau$. However, the bottom panels depict the logarithmic correction of the SDWs in the weak gap regime. In addition to the decreasing behavior with respect to the parameters $x$ and $\tau$, there is an oscillatory trend with respect to $x$. Furthermore, compared to the undisturbed state, the logarithmic correction enhances the y and z components of SDW correlations, along with the mixed components, and additionally contributes to the further renormalization of its power-law part due to interactions.

In the second case, we assume the system is not at the separatrix. So we use the solutions (\ref{yintsolution}) and (\ref{alphasolution}) and obtain,
\begin{align}
L_1(r)&=[\cosh(A l_r)-\frac{8\mathcal{Y}_{int}\sinh(A l_r)}{A}]^{-4},\label{tri1}\\
L_{2}(r)&=\exp[\frac{\cos(\alpha(l_r)x)}{\cos(\alpha(0)x)}-1].\label{tri3}
\end{align}
In this case, there are two different qualitative behaviors for $L_1$ and $L_{2}$ depending on the scale $l_r=log(\frac{r}{a_0})$. If $A l_r\ll1$ and $B l_r\ll1$, one can expand (\ref{tri1}) and (\ref{tri3}) as well as the trigonometric functions in Eq. (\ref{alphasolution}). To the lowest order in $A$ and $B$, we find that for $L_1$, the results on the separatrix ($A=0$) can be recovered. But for $L_{2}$, the result on the separatrix returns with a negative exponential coefficient for $\alpha$ where
\begin{equation}
    \alpha(l_r)=e^{\frac{-B^2}{\pi^2log(\alpha(0))}}e^{\frac{-1}{\pi log(\frac{r}{a_0})}}.
\end{equation}
This means that if the length scale is short enough the gap of the system regulars with the interactions so that the dimensionless coefficient of the gap is in a linear relation with the interactions; $\mathcal{Y}=8\mathcal{Y}_{int}$. However, the limit of $B l_r\ll1$ does not end at $B=0$. Instead, a deviation is observed. Moreover, in the length scale $log(\frac{r}{a_0})=\frac{1}{A}$, there is a crossover from a regime with linear relation between the gap and interactions to a regime with non-linear behavior which is renormalized by the length scale $log(\frac{r}{a_0})=\frac{1}{B}$. Now, if $A l_r\gg1$ and $B l_r\gg1$, by expanding $L_1(r)$ and $\alpha(l_r)$, we get the following expression for the correlation functions,
\begin{align}
\mathcal{R}^{log,yy}_{SDW}(r)=\mathcal{R}^{log,zz}_{SDW}(r)=\frac{8\cos(\alpha(0)x)}{(\pi a_0)^2}(\frac{a_0}{r})^{16\mathcal{Y}_{int}+4+4A}\nonumber\\
    \times (1-\frac{8\mathcal{Y}_{int}}{A})^{-4}\exp[\frac{\cos(e^{\frac{-i\sqrt{2}B}{\pi}}x)}{\cos(\alpha(0)x)}-1],
    \label{logcorrecty}
\end{align}
\begin{align}
\mathcal{R}^{log,zy}_{SDW}(r)=-\mathcal{R}^{log,yz}_{SDW}(r)=\frac{8\sin(\alpha(0)x)}{(\pi a_0)^2}(\frac{a_0}{r})^{16\mathcal{Y}_{int}+4+4A}\nonumber\\
    \times (1-\frac{8\mathcal{Y}_{int}}{A})^{-4}\exp[\frac{\cos(e^{\frac{-i\sqrt{2}B}{\pi}}x)}{\cos(\alpha(0)x)}-1].
    \label{logcorrectyz}
\end{align}
This situation is similar to a system with massless mode but with the renormalized Luttinger parameter $2K^*=16\mathcal{Y}_{int}+4+4A$. For correlation lengths large enough, where $\mathcal{Y}\neq8\mathcal{Y}_{int}$ and $\mathcal{Y}\neq\pi log(\alpha)$, there is no logarithmic correction. Instead, there are power-law corrections that become negligible as $r \rightarrow \infty$. 
This implies that in the case of the irrelevant gap, the y and z components of the SDW behave similarly to the normal Luttinger liquid but with a different scaling power compared to the non-perturbative mode, manifesting the additional effect of interactions in the power. 

 \subsubsection{Correction of residual correlations}

In this section, we calculate the non-exponential correlation functions by expanding the perturbation (\ref{SGor}) and obtain their corrections by applying second-order perturbation theory. Note that due to the exclusive reliance on the sine-Gordon form of operators (\ref{ySDWOp}) and (\ref{zSDWOp}), the y, z, and yz components of the SDW are constrained to undergo solely logarithmic corrections, which were discussed in the previous sections. Since the correlations of the mixed components xy and xz are zero in the unperturbed case, therefore they cannot have any perturbative corrections. On the other hand, given that the operators associated with charge and the x component of the SDW, i.e., Eqs. (\ref{CDWOp}) and  (\ref{xSDWOp}), take the form of gradients of the bosonic fields, it is necessary to investigate the residual corrections affecting both the CDW and the x component of the SDW correlation functions. 

We start with the unperturbed Matsubara correlation functions of the CDW and the x component of SDW, given by
\begin{align}
\mathcal{R}_{CDW}(i\omega_n)&=\int d\tau e^{i\omega_n}\langle T_{\tau}\mathcal{O}_{CDW}(\tau)\mathcal{O}_{CDW}(0)\rangle,\\
\mathcal{R}_{SDW}^{xx}(i\omega_n)&=\int d\tau e^{i\omega_n}\langle T_{\tau}\mathcal{O}_{SDW}^x(\tau)\mathcal{O}_{SDW}^x(0)\rangle.
\end{align}
Using translational invariance, we find the full expressions for the Matsubara correlation functions in the frequency-momentum space as,
\begin{align}
\mathcal{R}_{CDW}(k,i\omega_n)&=\frac{1}{4\pi}\int dr e^{i(\omega_n\tau-kx)}\langle T_{\tau}\partial_x\tilde{\Phi}(r)\partial_x\tilde{\Phi}(0)\rangle\\
\mathcal{R}_{SDW}^{xx}(k,i\omega_n)&=\frac{1}{4\pi}\int dr  e^{i(\omega_n\tau-kx)}\langle T_{\tau}\partial_x\Theta(r)\partial_x\Theta(0)\rangle
\end{align}
The frequency dependence of the unperturbed correlation functions can be calculated by taking the analytic continuation $i\omega_n\rightarrow \omega+i\eta^+$ and integrating over k, yielding [see Appendix \ref{Fouriertransform}], 
\begin{align}
\mathcal{R}_{CDW}(\omega)&=-\frac{i}{8}K\omega e^{-\frac{\omega a_0}{v}} Sign(\omega),\\
\mathcal{R}_{SDW}^{xx}(\omega)&=-\frac{i}{8K}\omega e^{-\frac{\omega a_0}{v}} Sign(\omega).
\end{align}
Here, the dependence on the Luttinger parameter manifests itself in the coefficients. Similar cases with different coefficient dependencies were observed in \cite{Citro-Giamarchi}.

When the gap term is irrelevant in the RG sense (Eq. \ref{flowgap}), it contributes to the residual correlations up to the second order in the gap coefficient, resulting in the residual corrections as:
 \begin{align}
     \mathcal{C}_{CDW}(r)=\frac{\mathcal{Y}^2(l)}{64\pi^3 a_0^4}\sum_{\epsilon=\pm1}e^{-i\alpha\epsilon(x_3-x_4)}\nonumber\\\times\langle T_{\tau}\partial_{x}\tilde{\Phi}(r)\partial_{x}\tilde{\Phi}(0)e^{i\sqrt{4\pi}\epsilon \tilde{\Phi}(r_3)}e^{-i\sqrt{4\pi}\epsilon \tilde{\Phi}(r_4)}\rangle_{\mathcal{H}_{hel}},
 \end{align}
for the CDW and 
  \begin{align}
     \mathcal{C}^{xx}_{SDW}(r)=\frac{\mathcal{Y}^2(l)}{64\pi^3 a_0^4}\sum_{\epsilon=\pm1}e^{-i\alpha\epsilon(x_3-x_4)}\nonumber\\\times\langle T_{\tau}\partial_{x}\Theta(r)\partial_{x}\Theta(0)e^{i\sqrt{4\pi}\epsilon \tilde{\Phi}(r_3)}e^{-i\sqrt{4\pi}\epsilon \tilde{\Phi}(r_4)}\rangle_{\mathcal{H}_{hel}},
 \end{align}
for the x component of the SDW. These averages can be estimated, respectively, from the correlators
 \begin{equation}
     \sum_{\epsilon=\pm1}\langle T_{\tau}e^{i\lambda\partial_{x}\tilde{\Phi}(r)}e^{i\mu\partial_{x}\tilde{\Phi}(0)}e^{i\sqrt{4\pi}\epsilon \tilde{\Phi}(r_3)}e^{-i\sqrt{4\pi}\epsilon \tilde{\Phi}(r_4)}\rangle_{\mathcal{H}_{hel}},
 \end{equation}
and
 \begin{equation}
     \sum_{\epsilon=\pm1}\langle T_{\tau}e^{i\lambda\partial_{x}\Theta(r)}e^{i\mu\partial_{x}\Theta(0)}e^{i\sqrt{4\pi}\epsilon \tilde{\Phi}(r_3)}e^{-i\sqrt{4\pi}\epsilon \tilde{\Phi}(r_4)}\rangle_{\mathcal{H}_{hel}},
 \end{equation}
by taking the first derivative with respect to $\lambda$ and $\mu$ in the limit $\lambda,\mu\rightarrow0$ and using the generalized Debye-Waller's relation,
\begin{align}
     \langle &T_{\tau}\Pi_je^{i(A_jf(\tilde{\Phi}(r_j))+B_jf(\Theta(r_j)))}\rangle\nonumber\\&=e^{-\frac{1}{2}\langle T_{\tau}[\sum_j (A_j f(\tilde{\Phi}(r_j))+B_jf(\Theta(r_j)))]^2\rangle}.
\end{align}
Here, $f(\tilde{\Phi}(r_j)), f(\Theta(r_j))=\partial_{x}\tilde{\Phi}(r_j), \partial_{x}\Theta(r_j)$ or $\tilde{\Phi}(r_j), \Theta(r_j)$. So, the corrections for the CDW and the x component of the SDW recast as
  \begin{widetext}
      \begin{equation}
     \mathcal{C}_{CDW}(r)=\frac{\mathcal{Y}^2(l)}{4\pi^2 a_0^4}\int d^2r_3 d^2r_4 e^{-2\pi \langle T_{\tau}[\tilde{\Phi}(r_3)-\tilde{\Phi}(r_4)]^2\rangle}\\\langle T_{\tau}[\partial_x \tilde{\Phi}(r)\tilde{\Phi}(r_3)]\rangle\langle T_{\tau}[\partial_x \tilde{\Phi}(0)\tilde{\Phi}(r_3)-\partial_x \tilde{\Phi}(0)\tilde{\Phi}(r_4)]\rangle \cos[\alpha (x_3-x_4)],
     \label{1correlationresidual}
 \end{equation}
 \begin{equation}
      \mathcal{C}^{xx}_{SDW}(r)=\frac{\mathcal{Y}^2(l)}{4\pi^2 a_0^4}\int d^2r_3 d^2r_4 e^{-2\pi \langle T_{\tau}[\tilde{\Phi}(r_3)-\tilde{\Phi}(r_4)]^2\rangle}\\\langle T_{\tau}[\partial_x \Theta(r)\tilde{\Phi}(r_3)]\rangle\langle T_{\tau}[\partial_x \Theta(0)\tilde{\Phi}(r_3)-\partial_x \Theta(0)\tilde{\Phi}(r_4)]\rangle \cos[\alpha (x_3-x_4)].
      \label{2correlationresidual}
 \end{equation}
  \end{widetext}
The correlators (\ref{1correlationresidual}) and (\ref{2correlationresidual}) contain the Hartree and Fock parts. As shown in Appendix \ref{correctgap}, the Hartree term doesn't have any contribution to the correction for $K>2$, but the Fock term has a remarkable contribution. Overall, the frequency dependence of the residual correlations, including the unperturbed terms and the corrections, in the presence of the irrelevant gap can be obtained as [see Appendix \ref{correctgap}]
\begin{align}
&\mathcal{R}^{Residual}_{CDW}(\omega)\approx\omega +\mathcal{Y}^2 \omega^{2(K-2)},\\
&\mathcal{R}^{xx, Residual}_{SDW}(\omega)\approx\omega+\mathcal{Y}^2 \omega^{2K-3}.
\end{align}
In the above relations, the second terms originate from the corrections to the residual correlation. The x component of the SDW and the CDW phase undergo an unconventional Luttinger liquid damping. For the CDW when $K>\frac{5}{2}$, the correction term becomes subdominant concerning $\omega$ as $\omega \rightarrow0$. While the correction for the x component of the SDW adds a subdominant contribution when $K>2$. Therfore, as the correction for the x component of SDW becomes negligible when $K>2$, the CDW correction remains significant and persists up to $K=\frac{5}{2}$. This indicates that the CDW phase, influenced by the mixed helical state, competes with the y and z components of the SDW at the gap relevancy frontier, while the helical x component of the SDW remains unaffected.

\section{Charge conductivity}\label{s5}
One of the quantities that characterizes the system's features is charge conductivity, which describes the linear current response to an external electric field. When the helical system is subjected to an exchange field, the band edge at $k=0$ bends, causing a change in the velocity of carriers near the edge, and leading to a mixing of spin modes between right movers and left movers. This properties may affect the conductivity. So, in this section, we calculate the charge conductivity in the presence of the gap term using the Memory function method \cite{umklapp-Giamarchi}. Additionally, to improve the validity of our calculations across a wide range of temperatures, we integrate the RG analysis with the results obtained from the Memory function. 

Before going further, it should be noted also that due to the significant difference between temperatures up to room temperature and the energy scales of Fermi levels in typical metals and topological insulators, the level-broadening effects induced by temperature are minimal in our context. More precisely, at these low temperatures, the thermal energy $k_BT$ is smaller than the Fermi energy $\epsilon_F$, leading to negligible thermal smearing of the Fermi surface. As a result, excitations near the Fermi level are not significantly affected by temperature-induced broadening effects. This justifies our approach of analyzing the system's behavior at the Fermi level without accounting for such broadening. 

Based on the Kubo formula for conductivity \cite{Mahan}, one can write the conductivity as \cite{umklapp-Giamarchi}
\begin{equation}
\sigma({\omega,T})=\frac{ivK}{\pi} \frac{1}{\omega+M(\omega,T)},
\label{Condct}
\end{equation}
where 
\begin{equation}
    M(\omega,T)=\frac{\omega \chi({\omega,T})}{\chi({0,T})-\chi({\omega,T})},
    \label{Memo}
\end{equation}
is the Memory function with $\chi(\omega,T)$ being the current-current correlation function. As presented in Appendix \ref{Memoryfunction}, we find the charge conductivity, at $\omega=0$,  as
\begin{align}
&\sigma(\omega=0,T)\propto\Delta_z^{-2}(\frac{2\pi a_0T}{v})^{3-2K}\nonumber\\
&\times \mathbf{B}(\frac{K}{2}-i\frac{v\alpha}{4\pi T},1-K)^{-1}\mathbf{B}(\frac{K}{2}+i\frac{v\alpha}{4\pi T},1-K)^{-1}\nonumber\\
&\times(\cot(\pi\frac{K}{2}+i\frac{v\alpha}{4T})+\cot(\pi\frac{K}{2}-i\frac{v\alpha}{4T}))^{-1}.
\label{conduct0}
\end{align}
In addition, at high temperature,  $T\gg\Delta_z, v\alpha,\omega$, we obtain the approximated expression for $\sigma(\omega=0,T)$ as 
\begin{equation}
\sigma_{app}(\omega=0,T)\propto\Delta_z^{-2}T^{3-2K}\frac{1}{D}(1+C(\frac{v\alpha}{4\pi T})^2),
\label{conductivityhighT0}
\end{equation}
where $C$ and $D$ are the dimensionless coefficients depending on the Luttinger parameter $K$ being given in Appendix \ref{Memoryfunction}. One can explicitly see that the temperature dependence of conductivity behaves as power-law with exponent controlled by interaction. Also, the main and approximated expressions of the conductivity match well for $T > (\Delta_z, v\alpha)$ (Not shown)

It is obvious that the unusual power-law behavior obtained for the charge conductivity is valid only in the high-temperature limit. To extend the description of qualitative behaviors across a wide temperature range with $T\gtrsim\Delta_z$, we amplify these findings through RG analysis. As for the RG flow of the gap in Eq. (\ref{flowgap}), this parameter scales as
\begin{equation}
\mathcal{Y}(l)=\mathcal{Y}(0) e^{(2-K(l))l}.
\end{equation}
Here, we assumed that 
$K(l)$ varies slowly with $l$. In this case, the temperature serves as the energy cutoff, so one must renormalize to a cutoff approximately of the order of the thermal length $L_{T}\sim{\frac{v}{T}}$, corresponding to $e^{l^*}\simeq e^{L_T}\sim{\frac{L_T}{a_0}}$ to access a relevant solution. By stopping the scaling treatment at $L_T$, we can utilize the values $\Delta_z(L_T),K(L_T),\alpha(L_T),v(L_T)$ as the temperature-dependent quantities in the Memory function as well as the conductivity instead of the bare parameters $\Delta_z$, $K$, $\alpha$, $v$. So, we get the RG-improved expression for the conductivity as
\begin{widetext}
\begin{align}
\sigma(L_T)=&\frac{v(L_T)}{2\pi^6T}\mathcal{Y}(L_T)^{-2}\sin(\pi K(L_T))^{-1}\mathbf{B}(\frac{K(L_T)}{2}-i\frac{v(L_T)\alpha(L_T)}{4\pi T},1-K(L_T))^{-1}\nonumber\\\times&\mathbf{B}(\frac{K(L_T)}{2}+i\frac{v(L_T)\alpha(L_T)}{4\pi T},1-K(L_T))^{-1}(\cot(\pi\frac{K(L_T)}{2}+i\frac{v(L_T)\alpha(L_T)}{4T})+\cot(\pi\frac{K(L_T)}{2}-i\frac{v(L_T)\alpha(L_T)}{4T}))^{-1}.
\label{sigmaLT}
\end{align}
\end{widetext}
Note that at high temperatures, the gap coefficient will act as a constant value and will not undergo re-scaling. 

At temperatures below the gap, $T<(v\alpha,\Delta_z)$, there are two cases which will be investigated below. When $T<\Delta_z<v\alpha$, the RG-improved conductivity (\ref{sigmaLT}) is expected to show a correct behavior. At very low temperatures, $T\ll v\alpha$, the conductivity can be approximated as
\begin{align}
\sigma\approx&16^{K(L_T)-1}\pi^{2K(L_T)-8}\mathcal{Y}(L_T)^{-2}(\frac{v(L_T)\alpha(L_T)}{T})^{1-K(L_T)}\nonumber\\
    \times&e^{\frac{v(L_T)\alpha(L_T)}{2T}}\frac{v(L_T)}{4T}\sin(\pi K(L_T))^{-2}\Gamma(1-K(L_T))^{-2}.
\end{align}
In this case the Fermi level lies above the gap providing charge carriers in the conduction band. So, the conductivity increases exponentially as the temperature decreases. 
 
But when $T<v\alpha<\Delta_z$, the system enters the strong gap phase. In the strong gap phase, the gap becomes relevant at $K<2$ under the RG, leading to electron localization and a decrease in the number of charge carriers with strong repulsive interactions. Subsequently, indeed, one may anticipate an exponential decaying behavior of the conductivity as a function of the electronic gap. In this regime, the smallest length scale at temperatures above zero is determined by the gap, rendering the Memory function approach invalid. However, as indicated by previous studies \cite{10visooriGiama2020}, the Memory function approach can still be utilized in the regime where the cosine term becomes relevant before $\mathcal{Y}(l) \sim 1$. So the conductivity can be written as $\sigma(T)=\sigma(\Delta_Z)e^{-\Delta_Z/T}$, where $\Delta_Z=\Delta\mathcal{Y}^{\frac{1}{2-2K}}$. Here, $\Delta=\frac{\hbar v_F}{a_0}$ represents the bulk gap, with $a_0$ being the transverse decay length, and $\sigma(\Delta_Z)$ is a constant coefficient determined by the Memory function expression \cite{10visooriGiama2020, 11Hsu-nuclear-chargetransp}. This exponential decay behavior was attributed to the influence of Instantons, which are employed to study the tunneling events across the gap \cite{instanton1, instanton2}.

\begin{figure}[t!]
    \centering
    \includegraphics[width=8cm]{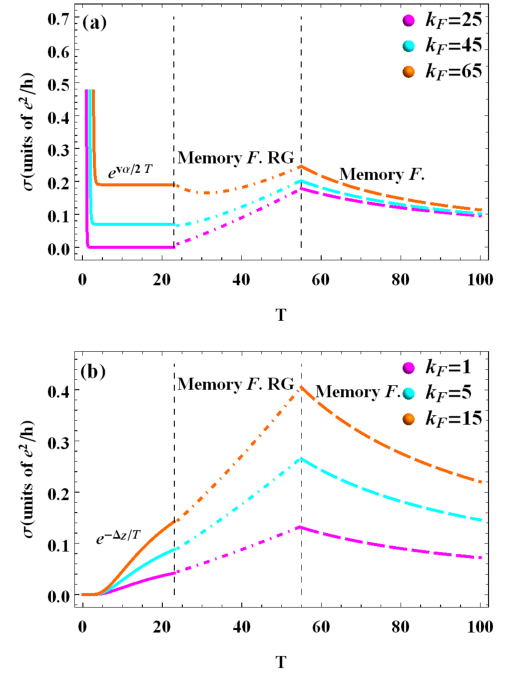}
   \caption{The charge conductivity as a function of temperature in the three temperature regions determined by different procedures for (a) large and (b) small Fermi momenta, with $\Delta_z=20$ and K=0.85. The dashed lines correspond to initial Memory function, the dot-dashed line to the RG-improved Memory function and the solid lines to the exponential behavior. To ensure continuity between the three regions at the boundaries, we adjusted the amplitude of conductivity in each region while preserving the overall conductivity behavior. Therefore, we matched the values of the charge conductivity at the transition temperatures by connecting the expressions derived from the three methods.}
    \label{conductivity-3part}
\end{figure}

For numerical evaluation, we take $v_F=1$ and the lattice constant as the length unit. The overall behavior of the conductivity as a function of temperature, including the three temperature regions with corresponding equations discussed above, is illustrated in Fig. \ref{conductivity-3part}. Panels (a) and (b) depict scenarios for large and small Fermi momenta, respectively. In both panels, transitioning from high temperatures to temperatures on the order of the gap, conductivity exhibits an ascending-descending trend. At elevated temperatures, decreasing temperature leads to an increase in conductivity. However, upon reaching temperatures comparable to the gap, a downward trend emerges in conductivity. This transition corresponds to moving from the crossover regime to the strong gap regime, where the growth of the gap coefficient $\mathcal{Y}(L_T)^2$ begins to impact conductivity. 
Moreover, as illustrated in Fig. \ref{conductivity-3part}(a), at very low temperatures, the conductivity increases due to the large Fermi momentum, corresponding to a higher Fermi level compared to the energy gap. In contrast, at the same temperatures, as depicted in Fig. \ref{conductivity-3part}(b), for small Fermi momentum causing the Fermi level to reside below the energy gap, the conductivity decreases. Moreover, from both panels, it is evident that increasing the Fermi momentum, and consequently, raising the Fermi level relative to the gap, leads to the enhancement of conductivity regardless of the temperature regions. Remarkably, these results hold true even in the case of $\alpha=0$. 

\begin{figure}[t!]
\centering
\includegraphics[width=8cm]{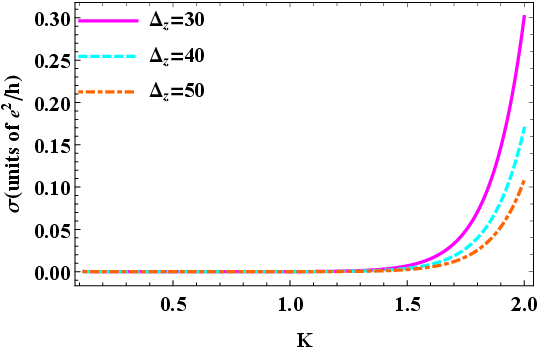}
\caption{(Color online)  Dependence of conductivity on the Luttinger parameter $K$ for various values of the gap coefficient at $k_F=30$ and $T=30$.}
\label{fig5}
\end{figure}

In addition, we have evaluated the conductivity versus Luttinger parameter $K$ by using the RG-improved conductivity, shown in Fig. \ref{fig5}, with different values of $\Delta_z$. As expected, the conductivity decreases for larger $\Delta_z$. We observe that in the presence of strong repulsive interactions, the wire is deeply embedded in the relevant regime ($K\ll1$), causing the gap to have a significant effect in reducing the conductivity, resulting in a notable drop.

\section {Summary and Conclusions} \label{s6}

In conclusion, our study investigate the interplay between electron-electron interactions and the Zeeman field in one-dimensional helical states. By employing a combination of theoretical tools including bosonization, renormalization group analysis, and Memory function technique, we have uncovered rich phenomena governing the low-energy properties of these systems. We identified three distinct regimes in the phase diagram: the strong gap, crossover, and weak gap regimes. In the strongly interacting regime, the Zeeman-generated gap becomes relevant, fundamentally altering the behavior of charge and spin correlations. 

Firstly, our investigation into the SDW correlations has revealed intriguing insights into their response to varying interaction strengths and Zeeman field orientations. We find that in the absence of significant Zeeman field effects, SDW correlations along different axes exhibit similar behavior, indicative of a disturbed Luttinger liquid regime. However, the introduction of a relevant Zeeman gap parameter leads to the dominance of certain SDW components, particularly those aligned with the field direction. This phenomenon highlights the critical role of the Zeeman field in stabilizing and modulating spin correlations in the system.

Furthermore, our analysis of CDW correlations elucidates the impact of the gap parameter on charge ordering phenomena. We observe that the emergence of a gap due to the mixing of helical states suppresses CDW correlations, effectively freezing their time-space evolution. This effect is particularly pronounced in the strong gap regime, where the regularization of the field leads to the disappearance of charge density wave features. Additionally, we find that the CDW correlations exhibit non-trivial behavior near the gap boundary, with logarithmic corrections enhancing certain phases while suppressing others. We observe the interactions play an effective role in phase transition of our desired system.

Furthermore, our analysis of charge conductivity revealed intriguing temperature dependencies, exhibiting non-uniform behaviors in both the strong and weak gap parameter regions. Above the gap and at high temperatures, we observed an increasing trend in conductivity with decreasing temperature, while near-gap temperatures witnessed interaction-controlled power-law decay. At temperatures below the gap, the presence of a strong gap led to an exponential decrease in conductivity, highlighting the profound influence of the gap parameter on charge transport properties. 

Helical states as edges of two-dimensional topological insulators (2DTIs), unlike their one-dimensional counterparts, have a potential in a wide range of applications from spintronics to quantum computing \cite{TI,Ts1}. Also, dissipationless transport which is ideal for low power consumption devices. On the other hand, our findings enhance our understanding of the transport properties of such systems and provide insights into manipulating quantum states. Potential applications include the development of low power consumption spintronic devices, advancements in quantum computing, and the creation of novel materials with tailored transport properties. Future research could explore optimizing these interactions for specific technological applications or investigating similar effects in other topological materials.

While our study is primarily theoretical, it is worth noting that experimental realization of systems hosting both strong interactions and strong spin-orbit coupling is feasible \cite{strong-eeexperiment}. Furthermore, investigating the charge transport properties \cite{e.transport-review}  offers a prosperous prob to extract experimental signatures from helical Luttinger liquids \cite{SOnanowire2,experiment-helicaltransport,2Sooriintro}.

\section*{Code availability} 
The codes are available at \cite{Code}.

\section*{Acknowledgement}
We are grateful to T. Giamarchi and R. Citro for valuable discussions.

\appendix
\begin{widetext}
   \section {Logarithmic corrections in the  weak gap regime} \label{logcorrect}
In this section, we investigate the effects caused by the gap on the correlation functions using the procedure known as logarithmic correction \cite{Giamarchi}. For example, the correction for the correlation of the y component of the SDW can be obtained by computing the correlators like
\begin{equation}
    \sum_{\epsilon=\pm1}\int dx_3 d\tau_3 dx_4 d\tau_4\langle e^{-\alpha\epsilon(x_3-x_4)}e^{\alpha(x_1-x_2)}e^{i\sqrt{4\pi}\tilde{\Phi}(x_1,\tau_1)}e^{-i\sqrt{4\pi}\tilde{\Phi}(x_2,\tau_2)}e^{i\sqrt{4\pi}\epsilon\tilde{\Phi}(x_3,\tau_3)}e^{-i\sqrt{4\pi}\epsilon\tilde{\Phi}(x_4,\tau_4)}\rangle_{\mathcal{H}_{hel},con.},
\end{equation}
where con. implies the connected correlation. After some calculations, we arrive at
\begin{equation}
    \mathcal{R}^{log,yy}_{SDW}(r,r^{\prime})=\frac{1}{2(\pi a_0)^2}\cos(\alpha x)e^{-2KF_1(r)}(\frac{a_0}{r})^{32\mathcal{Y}_{int}}L_1(r)L_2(r)\mathcal{S},
    \label{logcorrectyy}
\end{equation}
where
\begin{equation}
\mathcal{S}=\exp[-2\cos(2\theta_r)J_2(\alpha a_0)\int_{0}^{l_r}\mathcal{Y}^2 dl].
\label{S0}
\end{equation}
Here, $F_1(r)$, $L_1(r)$, and $L_2(r)$ are introduced in Eqs. (\ref{F1}), (\ref{L1}), and (\ref{L2}), respectively. In the non-perturbative Hamiltonian, $t$ is zero. But when the Zeeman field turns on, we see in the RG analyze, the perturbative expansion contains the expression $\cos(2\theta_r)$ steamed from the parameter $\alpha$. This requires that Eq. (\ref{F1}) is considered fully. Taking $K=2-8\mathcal{Y}_{int}$, the expression (\ref{logcorrectyy}) can be read as
\begin{equation}
\mathcal{R}^{log,yy}_{SDW}(r,r^{\prime})=\frac{1}{2(\pi a_0)^2}\cos(\alpha x)(\frac{a_0}{r})^{4+16\mathcal{Y}_{int}}e^{-2t\cos(2\theta_r)}L_1(r)L_2(r)\mathcal{S}.
\label{2logcorrectyy}
\end{equation}
Plugging the flow (\ref{tyRG}) to (\ref{S0}) yields
\begin{equation}
\mathcal{S}=exp[-4\int_{0}^{l_r}\frac{dt(l)}{dl}\cos(2\theta_r)dl]=\exp[4\cos(2\theta_r)t(0)]L_3(r),
\label{S1}
\end{equation}
with $L_3(r)=e^{-4\cos(2\theta_r)t(l_r)}$. Substituting Eq. (\ref{S1}) into (\ref{2logcorrectyy}) gets Eq. (\ref{finalLogCorr}) of the main text.

On the other hand, combining the flow Eqs. (\ref{yintRG}) and (\ref{tyRG}), we get 
\begin{equation}
    \frac{dt(l)}{dl}=4\frac{d\mathcal{Y}_{int}(l)}{dl}J_2(\alpha(l)a_0),
\end{equation}
leading to the solution
\begin{equation}
    t(l)=4 \mathcal{Y}_{int}J_2(\alpha(l)a_0).
\end{equation}
According to this equation and the result obtained from $\mathcal{Y}_{int}$ in $r\rightarrow \infty$, one can obtain $t(l_r)\sim \frac{-4J_2(\alpha(l)a_0)}{l_r}$. In such limit, one gets 
\begin{equation}
L_3(r)=e^{16\cos(2\theta_r)log(\frac{a_0}{r})J_2(\alpha(l)a_0)}.
\end{equation}
As $r\rightarrow \infty$, $\cos(2\theta_r)=2\frac{x}{r}^2-1=-1$, so $L_3(r)$ reduces as 
\begin{equation}
     L_3(r)=(\frac{a_0}{r})^{-16J_2(\alpha(l)a_0)}.
\end{equation}
But, when the cutoff $a_0$ tends to zero, one can consider $J_2(\alpha(l)a_0)\ll1$. Thus, we incorprate $L_3(r)\approx1$ in the expressions for the correction in Eqs. (\ref{0logcorrectyy})-(\ref{logcorrectyz}).

\end{widetext}

\section {Fourier transform of the non-perturbative residual correlation} \label{Fouriertransform}

The Fourier-transformed correlator in Eq. (\ref{F2}) is given by
\begin{align}
&F_2(k,i\omega_n)=\nonumber\\
&\int_{-\infty}^{\infty}dx\int_{-\infty}^{\infty}d\tau \frac{(v\tau sign(\tau)+a_0)^{2}-x^2}{2((v\tau sign(\tau)+a_0)^{2}+x^2)^2}e^{i(\omega_n\tau-kx)}.
\end{align}
Integrating over $x$ yields:
\begin{equation}
F_2(k,i\omega_n)=\frac{\pi}{2}\int_{-\infty}^{\infty}d\tau e^{i(\omega_n-kv)|\tau|}e^{-ka_0}k,
\end{equation}
and further integration over 
$\tau$ gives,
\begin{equation}
F_2(k,i\omega_n)=\frac{\pi}{2} k e^{-ka_0}(\frac{1}{i\omega_n+vk}-\frac{1}{i\omega_n-vk}).
\label{F2k}
\end{equation}
Substituting $i\omega_n$ with $\omega+i\eta^+$, we can perform the analytic continuation of equation (\ref{F2k}) as follows:
\begin{equation}
    F_2(k,\omega)=-i\frac{\pi^2}{2} k e^{-ka_0}[\delta(\omega-vk)-\delta(\omega+vk)].
\end{equation}
Here, we utilize the identity,
\begin{equation}
    \frac{1}{x+i\eta^+}=P(\frac{1}{x})-i\pi \delta(x).
\end{equation}
To examine the frequency dependence of the correlation function, we integrate over 
$k$ in the expression:
\begin{equation}
    F_2(k,\omega)=-i\frac{\pi^2}{2}\int_0^{\infty}dk k e^{-ka_0}[\delta(\omega-vk)-\delta(\omega+vk)].
\end{equation}
This integration yields,
\begin{equation}
   F_2(\omega)=-i\frac{\pi^2}{2} \omega e^{-\frac{\omega a_0}{v}}Sign(\omega).
\end{equation}

\section {Correction of residual correlations by gap operator} \label{correctgap}
\begin{widetext}

The correction for the charge density wave (CDW) is given by
\begin{equation}
\mathcal{C}_{CDW}(r)=\frac{\mathcal{Y}^2(l)}{4\pi^2 a_0^4}\int d^2r_3 d^2r_4 e^{-2\pi \langle T_{\tau}[\tilde{\Phi}(r_3)-\tilde{\Phi}(r_4)]^2\rangle}\\\langle T_{\tau}[\partial_x \tilde{\Phi}(r)\tilde{\Phi}(r_3)]\rangle\langle T_{\tau}[\partial_x \tilde{\Phi}(0)\tilde{\Phi}(r_3)-\partial_x \tilde{\Phi}(0)\tilde{\Phi}(r_4)]\rangle \cos[\alpha (x_3-x_4)],
\label{CCDW}
\end{equation}
where the correlations are given by
\begin{equation}
\langle T_{\tau}[\partial_x \tilde{\Phi}(x,\tau)\tilde{\Phi}(x_3,\tau_3)]\rangle=\partial_xG(x-x_3,\tau-\tau_3)=-\frac{K}{2}\frac{x-x_3}{(x-x_3)^2+(v|\tau-\tau_3|+a_0)^2},
\end{equation}
\begin{equation}
\langle T_{\tau}[\partial_x \tilde{\Phi}(0,0)\tilde{\Phi}(x_3,\tau_3)]\rangle=\partial_xG(x_3,\tau_3)=\frac{K}{2}\frac{x_3}{x_3^2+(v|\tau_3|+a_0)^2},
\end{equation}
and similarly for the term with $3\rightarrow4$. Meanwhile,
\begin{equation}
     e^{-2\pi \langle T_{\tau}[\tilde{\Phi}(r_3)-\tilde{\Phi}(r_4)]^2\rangle}=(\frac{a_0^2}{(x_3-x_4)^2+(v|\tau_3-\tau_4|+a_0)^2})^K.
\end{equation}
  
Assuming $\tau_3-\tau_4=\tau^{\prime}$ and $x_3-x_4=x^{\prime}$, in Eq. \ref{CCDW}, one can factorize the term as follows,
\begin{multline}
\int\int dx^{\prime} d\tau^{\prime} (\frac{a_0^2}{(x^{\prime})^2+(v|\tau^{\prime}|+a_0)^2})^K\cos(\alpha x^{\prime})\\=\int \frac{dz}{(1+z^2)^K}\int d\tau^{\prime}\frac{a_0^{2K}}{(a_0+v\tau^{\prime})^{2K-1}}\cos(\alpha z(a_0+v\tau^{\prime}))
=\frac{2a_0}{v}I(K),
\end{multline}
where $I(K)$ is a function solely dependent on $K$ and doesn't exhibit any power-law decay for $K>2$. Diagnosing the convolution integral,
\begin{equation}
\int dx_3 \int d\tau_3 \partial_x G(x-x_3,\tau-\tau_3)\partial_x G(x_3,\tau_3)=\int \frac{dk}{2\pi}\int \frac{d\omega}{2\pi}k^2|G(k,\omega)|^2e^{i(kx-\omega\tau)},
\end{equation}
we obtain the succinct Hartree correction $\mathcal{C}^{Hartree}(x,\tau)$,
\begin{equation}
\mathcal{C}^{Hartree}(x,\tau)=\frac{2a_0}{v}I(K)\int \frac{dk}{2\pi}\int \frac{d\omega}{2\pi}k^2|G(k,\omega)|^2e^{i(kx-\omega\tau)}=\frac{2a_0}{v}I(K)\int \frac{dk^{\prime}}{2\pi}\int \frac{d\omega^{\prime}}{2\pi}k^{\prime 2}G(k^{\prime},\omega^{\prime})G(-k^{\prime},-\omega^{\prime}),
\end{equation}
where $G(k^{\prime},\omega^{\prime})=\frac{vK}{\omega^{\prime 2}+(vk^{\prime})^2}$.
Utilizing the identity:
\begin{equation}
    e^{iax}=\int dk \mathcal{F}[e^{iax}](k)e^{ikx}=\int dk \delta(k-a)e^{ikx},
\end{equation}
with
\begin{equation}
    \mathcal{F}[e^{iax}](k)=\int dx e^{iax} e^{-ikx}=\int dx e^{-i(k-a)x}=\delta(k-a),
\end{equation}
we observe that as $\omega^{\prime}$ tends to zero, the integral does not decay, allowing us to omit this part of the correction.

The contribution of the Fock correction can be expressed as
\begin{equation}
\int d3 \int d4(\frac{(a_0)^2}{(a_0+v|\tau_3-\tau_4|)^2+(x_3-x_4)^2})^K\cos(\alpha (x_3-x_4))\partial_x G(x-x_3,\tau-\tau_3)\partial_x G(x_4,\tau_4).
\end{equation}
Using the Fourier transform representation, we obtain,
\begin{multline}
A = \mathcal{F}\left[\frac{1}{(a_0 + v|\tau'|)^2 + x'^2}\cos(\alpha x')\right] = \int dx' \int d\tau' \frac{e^{-ikx' + i\omega \tau'}}{(a_0 + v|\tau'|)^2 + x'^2} \cos(\alpha x') \\
= \left[\sqrt{(k \mp \alpha)^2 + \frac{\omega^2}{v^2}} a_0\right]^{K-1} \frac{1}{2^K \Gamma(K+1)} K_{K-1} \left[\sqrt{(k \mp \alpha)^2 + \frac{\omega^2}{v^2}} a_0\right],
\end{multline}
where $\Gamma$ is the Euler-Gamma function and $K$ is the Bessel function of the second kind. By defining $k\mp \alpha=\frac{\delta k}{2}\rightarrow k$, we can simplify further to obtain
\begin{equation}
A = \left[\sqrt{k^2 + \frac{\omega^2}{v^2}} a_0 \right]^{K-1} \frac{1}{2^K \Gamma(K+1)} K_{K-1} \left[\sqrt{k^2 + \frac{\omega^2}{v^2}} a_0 \right].
\label{Fock}
\end{equation}
Expanding for small $k$ and $\omega$, and considering $K-1$ as non-integer, the first non-zero order dependent on $k$ and $\omega$ is $(k^2+\frac{\omega^2}{v^2})^{K-1}$, which is subordinate compared to $k^2+\frac{\omega^2}{v^2}$ when $K>2$. In the diagrammatic representation, Eq. (\ref{Fock}) (up to a factor $\mathcal{Y}^2$) represents the self-energy correction as follows:
\begin{equation}
\Sigma(k,\omega)\approx \Sigma(0,0)+C(k^2+\frac{\omega^2}{v^2})^{K-1},
\end{equation}
where we omit holomorphic terms of order $(k^2+\frac{\omega^2}{v^2})a_0^2$ and higher. Consequently, the Fock correction is obtained (with regard to Wick’s theorem) as:
\begin{align}
    &-\int \frac{d\omega^{\prime}}{2\pi}\int \frac{dk}{2\pi}k^2[\Sigma(k,\omega^{\prime})G(k,\omega^{\prime})G(-k,\omega-\omega^{\prime})+\Sigma(-k,\omega-\omega^{\prime})G(-k,\omega-\omega^{\prime})G(k,\omega^{\prime})]\nonumber\\
   & =-\mathcal{Y}^2 \int \frac{d\omega^{\prime}}{2\pi}\int \frac{dk}{2\pi}k^2(k^2+\frac{(\omega-\omega^{\prime})^2}{v^2}a_0^2)^{K-1}(k^2+\frac{(\omega-\omega^{\prime})^2}{v^2})^{-1}(k^2+\frac{\omega^{\prime 2}}{v^2})^{-1}.
\end{align}
By evaluating the powers, this expression exhibits a non-universal power-law behavior as $\omega^{2(K-2)}$. For $K>\frac{5}{2}$, it is subordinate compared to the non-perturbative term $\omega$ as $\omega\rightarrow0$. Therefore, when $K>\frac{5}{2}$ and the gap term is irrelevant, the charge density wave correlation behaves as:
\begin{equation}
    \mathcal{R}^{corrected}_{CDW}(\omega)\approx \omega+\mathcal{Y}^2 \omega^{2(K-2)}+O(\omega^{2(K-2)}).
\end{equation}
The procedure for the spin density wave (SDW) is identical to CDW.
\end{widetext}

\section {conductivity and the Memory function technique } \label{Memoryfunction}

\begin{widetext}
The numerator of the Memory function (\ref{Memo}) can be expressed as 
\begin{equation}
\omega\chi({\omega,T})=-\frac{1}{\omega}[{\langle F;F \rangle}_{\omega}-\langle F;F \rangle_{\omega=0}],
\end{equation}
with
\begin{equation}
{\langle F;F \rangle}_{\omega}=-i\int\limits_0^{\infty}dt e^{i\omega t}\langle[F(t),F(0)]\rangle,
\end{equation}
where $F=[\mathcal{H}_{\Delta},j(t)]$ with $j(t)=vK\partial_x\Theta$ being the charge current operator. When the coefficient of the sine-Gordon term is considered to be small enough, then $\chi({\omega\neq0})$ is small, so that the denominator of the Memory function can be approximated as ${\chi({0,T})-\chi({\omega,T})}\approx\chi({0,T})$. Since $F$ is proportional to the sine-Gordon term, as a perturbation term, up to the second order of the perturbation expansion, the correlator ${\langle F;F \rangle}_{\omega}$ can be calculated by the unperturbed states, i.e., ${\langle F;F \rangle}_{\omega}\approx {\langle F;F \rangle}_{\omega}^0$. Therefore, the Memory function can be approximated as 
\begin{equation}
M(\omega,T)\approx-\frac{\langle F;F \rangle_{\omega}^0-\langle F;F \rangle_{\omega=0}^0}{\omega \chi(0,T)}.
\label{Memory}
\end{equation} 
Evaluating  $\langle F;F \rangle_{\omega}^0$, one gets the numerator of (\ref{Memory}) in the low frequency $\omega$ as 
\begin{equation}
\begin{split}
\langle F;F \rangle_{\omega}^0-\langle F;F \rangle_{\omega=0}^0=&4i K\sin(\pi K)(\frac{2\pi a_0T}{v})^{2K-2}\\&\times\mathbf{B}(-i\frac{(\omega-v\alpha)}{4\pi T}+\frac{K}{2},1-K)\mathbf{B}(-i\frac{(\omega+v\alpha)}{4\pi T}+\frac{K}{2},1-K)(\cot(\pi\frac{K}{2}+i\frac{v\alpha}{4T})+\cot(\pi\frac{K}{2}-i\frac{v\alpha}{4T})),
\end{split}
\end{equation}
where $\mathbf{B}(x,y)$ denotes the Beta function. Also, for the denominator of (\ref{Memory}), we obtain,
\begin{equation}
  \chi(0,T)=-\frac{vK}{\pi}.
\end{equation}
Inserting these expressions into Eq. (\ref{Memory}) and then substituting into Eq. (\ref{Condct}), at $\omega=0$, yields
\begin{equation}
\sigma(\omega=0,T)\propto\Delta_z^{-2}(\frac{2\pi a_0T}{v})^{3-2K}\mathbf{B}(\frac{K}{2}-i\frac{v\alpha}{4\pi T},1-K)^{-1}\mathbf{B}(\frac{K}{2}+i\frac{v\alpha}{4\pi T},1-K)^{-1}(\cot(\pi\frac{K}{2}+i\frac{v\alpha}{4T})+\cot(\pi\frac{K}{2}-i\frac{v\alpha}{4T}))^{-1}.
\label{conduct}
\end{equation}
By considering the different scale regions, at high temperature, $T\gg\Delta_z, v\alpha,\omega$, the expression for $M(\omega=0,T)$ can be approximated as
\begin{align}
M_{app}(\omega=0,T)\approx&4iK\Delta_z^2\pi\sin(\pi K)(\frac{2\pi a_0T}{v})^{2K-2}\frac{1}{T}D(1-C(\frac{v\alpha}{4\pi T})^2).
\label{memoryfuncapp}
\end{align}
Then, substituting Eq. (\ref{memoryfuncapp}) into Eq. (\ref{Condct}), at $\omega=0$, the approximated conductivity takes the form
\begin{equation}
\sigma_{app}(\omega=0,T)\propto\Delta_z^{-2}T^{3-2K}\frac{1}{D}(1+C(\frac{v\alpha}{4\pi T})^2).
\label{conductivityhighT}
\end{equation}
Here, we have used the expansion $(1-\epsilon)^{-1}\approx1+\epsilon$ with $\epsilon\ll1$. Also, the dimensionless coefficients $C$ and $D$, depending on the Luttinger parameter $K$, are given by
\begin{align}
    C&=\pi^2 \csc(\frac{K\pi}{2})^2-4(P(1,1-\frac{K}{2})-P(1,\frac{K}{2}),\\
    D&=\frac{1}{\pi}2^{-1-2K}\cot(\frac{K\pi}{2})\Gamma(\frac{1}{2}-\frac{K}{2})^2\Gamma(\frac{K}{2})^2.
\end{align}
where $\Gamma$ and $P$ denote the Gamma and Poly Gamma functions, respectively.

\end{widetext}

\end{document}